\documentstyle[12pt]{article}

\def\be{\begin{equation}}
\def\ee{\end{equation}}
\def\bea{\begin{eqnarray}}    
\def\eea{\end{eqnarray}}

\newcommand{\na}{\nabla}

\begin{document}
\begin{flushright}
\end{flushright}
\pagestyle{plain}
\begin{center}
\LARGE{\bf Exact Solutions for Cosmological Models with a Scalar Field\\}
\vspace{.5cm}
\small
\vspace{1.5cm}
{\Large{\bf H. Motavali$^{1,2,3}$}}\footnote{e-mail address: motaval@theory.ipm.ac.ir},~~
{\Large{\bf M. Golshani$^{3}$}}\footnote{e-mail address: golshani@ihcs.ac.ir}  \\  \vspace{0.5cm}
\small{$^{1}$ Department of Physics, McMaster University, Hamilton, Ontario, Canada L8S~4M1\\
$^{2}$ Department of Physics, Amirkabir University of Technology, Tehran, Iran 15875-4413 \\
$^{3}$ Institute for Studies in Theoretical Physics and Mathematics,Tehran, Iran 19395-1795 \\}
\today 
\end{center}
\vspace{.5cm}
\small
\begin{abstract}
We consider the existence of a Noether symmetry in the scalar-tensor
theory of gravity in flat Friedman Robertson Walker (FRW) cosmology. 
The forms of coupling function $\omega(\phi)$ and generic potential
$V(\phi)$ are obtained by requiring the existence of a Noether symmetry
for such theory. We derive exact cosmological solutions of the field
equations from a point-like Lagrangian.
\end{abstract}
\section {Introduction}
Scalar-tensor theories of gravity have been widely used in recent years. 
In principle, they are related to many fundamental theories as
super-string theories, grand unified theories and quantum gravity.${}^{1,2}$
These theories allow the gravitational coupling to vary and act as a
dynamical field.${}^{3,4}$ In the Kaluza-Klein models this arises from a
variation of the size of the internal dimensions. In a cosmological
context such theories allow to look for dynamical answers for equations.

The general form of the extended gravitational actions in scalar-tensor
theories can be written in terms of the Brans-Dicke field $\phi$ and  
the strength of the coupling between the scalar field and gravity which 
is represented by the dynamical coupling function $\omega(\phi)$. 
Furthermore, a nontrivial potential $V(\phi)$ may be introduced in terms
of a scalar field which clearly affect the dynamics. Higher-order
gravitational theories are equivalent to a scalar-tensor theory when
$V(\phi)$ is non-zero but $\omega(\phi)=0$, which introduces
Yukawa-type corrections to the Newtonian potential.${}^{5}$ The case of
$V(\phi)=0$ ensures a strictly Newtonian weak field limit to lowest order.
When the coupling functional $\omega(\phi)$ is considered to be a constant
parameter, the scalar-tensor theory reduces to the Brans-Dicke theory.

In general, there is no unique method to determine the functional forms of
$\omega(\phi)$ and $V(\phi)$. In this paper, we first prove the existence of 
the Noether symmetry and then we use this symmetry to find these functionals. The 
resulting functional forms for $\omega(\phi)$ and $V(\phi)$ are not 
independent of each other. One may note that the Lagrangian in the 
action becomes point-like if we impose a flat Friedman Robertson 
Walker (FRW) metric.

The paper is organized as follows: the existence of the Noether symmetry
is proved in section 2. In section 3 the exact solutions of the
field equations are derived from a point-like Lagrangian for a quartic 
potential. The concluding remarks appear in the section 4.

\section {Noether Symmetry }
We consider the action of scalar-tensor theory in the form
\begin{eqnarray}
S[\Phi]={ \int d^4x {\sqrt {-g}}  \left \{ \Phi R+ \frac{
\omega(\Phi)}{\Phi} g^{\mu\nu} \na_\mu \Phi \na_\nu\Phi
-V(\Phi) \right \} }
\end{eqnarray}
where $ R$ is the scalar curvature, $\Phi$ denotes a real scalar
field, non-minimally coupled to gravity, and ${\omega(\Phi)}$ and ${V(\Phi)}$ 
represent the coupling function and generic potential, respectively.
The action (1) can be rewritten, with redefinition $\Phi=\phi^2$, as
\begin{eqnarray}
S[\phi]={ \int  d^4x  \sqrt{-g} \left \{\phi^2 R+4{ \bf \omega (\phi)} 
g^{\mu \nu} \na_\mu \phi\na_\nu\phi-V(\phi)\right\} }. 
\end{eqnarray}
In the cosmological case, when the space-time manifold is described by 
a flat FRW metric, the scalar curvature has the expression 
$R=-6({\ddot{a}}/{a}+{\dot{a}^2}/{a^2})$, in which $a$ is the scale
factor of the universe and the dot denotes the derivative with respect
to time. One can show simply that in this space-time manifold the
Lagrangian density related to the action (2) takes the point-like
form${}^{6,7}$
\begin{eqnarray}
{\cal{L}}= 12 a^2\dot{a} \phi \dot{\phi}+ 6a \dot{a}^2  \phi^2+a^3
(4\omega(\phi) \dot{\phi}^2-V(\phi)). 
\end{eqnarray}
The corresponding Euler-Lagrange equations are given by
\begin{eqnarray}
[\frac{\ddot{a}}{a}+\frac{1}{2}(\frac{\dot{a}}{a})^2]\phi^2+
[\ddot{\phi}+2(\frac{\dot{a}}{a})\dot{\phi}]\phi+\dot{\phi}^2-
(\omega(\phi) \dot{\phi}^2-\frac{1}{4}V(\phi))=0,
\end{eqnarray}
and
\begin{eqnarray}
\dot{\phi} \omega(\phi)+3\dot{\phi}
(\frac{\dot{a}}{a})\omega(\phi)+\frac{3}{2}[\frac{\ddot{a}}{a}+
(\frac{\dot{a}}{a})^2]\phi +\frac{1}{2}
(\omega^{'}(\phi) {\dot{\phi}}^2+\frac{1}{4}V^{'}(\phi))=0.
\end{eqnarray}
These equations are equivalent to the second order Einstein equation
and the Klein-Gordon equation, in the flat FRW space-time, respectively.

We examine Noether symmetry of the Lagrangian (3) in order to determine
the two unknown functions $\omega(\phi)$ and $V(\phi)$, as well as solving
its dynamical field equations.\\ The infinitesimal generator of the
Noether symmetry, namely the lift vector field $X$, is
\begin{eqnarray}
X=\alpha (\phi, a)\frac{\partial}{\partial a}+\beta
(\phi, a) \frac{\partial}
{\partial \phi}+\frac{d \alpha}{dt}\frac{\partial}{\partial \dot{a}} 
+\frac{d\beta}{dt}\frac{\partial}{\partial \dot{\phi}}.\nonumber
\end{eqnarray}
We try to determine $\alpha (\phi, a)$ and $\beta (\phi, a)$ by imposing 
the following condition${}^8$ 
\begin{eqnarray}
L_{X} {\cal L}=0,
\end{eqnarray}
where $L_{X}$ is the Lie derivative with respect to the vector field $X$. 
Condition (6) leads to the following differential equations 
\begin{eqnarray}
&{}&{3\alpha V(\phi)+a\beta V^{'}(\phi)=0}\\
&{}&{3\alpha\omega(\phi)+\beta
a\omega^{'}(\phi)+3\phi\frac{\partial\alpha}{\partial\phi}+
2a\omega(\phi)\frac{\partial \beta}{\partial\phi}=0}\\
&{}&{\alpha\phi^2+2\beta a \phi +2a\phi^2\frac{\partial \alpha}
{\partial a}+2a^2\phi \frac{\partial \beta}{\partial a}=0}\\
&{}&{2\phi\alpha+a\beta+a\phi\frac{\partial \alpha}{\partial a} 
+\phi^2\frac{\partial \alpha}{\partial\phi}+
a\phi\frac{\partial \beta}{\partial \phi}+\frac{2}{3}
a^2\omega(\phi)\frac{\partial \beta}{\partial a}=}0. 
\end{eqnarray}
From Eq. (7) we have
\begin{eqnarray}
\alpha=-a\beta U(\phi),
\end{eqnarray}
where we have used the definition
\begin{eqnarray}
U(\phi)=\frac{1}{3}\frac{V^{'}(\phi)}{V(\phi)}.
\end{eqnarray}
Substituting (11) into (8) and (9), we find that the variables $\phi$
and $a$ in $\beta(\phi,a)$ must separate as
\begin{eqnarray}
\beta(\phi,a)=f(\phi)a^n,
\end{eqnarray}
where the separation constant $n$ is given by
\begin{eqnarray} 
n=\frac{\frac{3}{2}\phi U(\phi)-1}{1-\phi U(\phi)},
\end{eqnarray}
and functional $f(\phi)$ has the following form
\begin{eqnarray}
f(\phi)=exp \left\{ \int { \frac{\omega(\phi) U(\phi)+\phi U^{'}(\phi)
-\frac{1}{3}\omega{'}(\phi) }{\frac{2}{3}\omega(\phi)-\phi U(\phi)} }d\phi
\right \}.
\end{eqnarray}
From the definition (12) and using (14) we obtain 
\begin{eqnarray}
V(\phi)=\lambda \phi^m,~~~~~~~~~~~m=\frac{3(n+1)}{(n+3/2)}
\end{eqnarray}
and
\begin{eqnarray}
U(\phi)=\frac{1}{3}m\phi^{-1}
\end{eqnarray}
where $\lambda$ is an arbitrary integration constant. 
Putting these results into Eq. (10) one has
\begin{eqnarray}
(m-3)\phi
\omega^{'}(\phi)+n\omega^2(\phi)-[m^2+m(1+4n)-6]\omega(\phi)+4(nm+3m-6)m=0.
\end{eqnarray}
This is a first-order differential equation for the dynamical coupling
function $\omega(\phi)$ with the solution
\begin{eqnarray}        
\omega (\phi)=k_1 [\frac{(\phi/\phi_0)^{k_2} -1}{(\phi/\phi_0)^{k_2} +1}]
+\omega_0,
\end{eqnarray}
where $\phi_0$ is an integration constant and $k_1$, $k_2$ and $\omega_0$
are defined as
\begin{eqnarray}
k_1=\frac{3}{(2n+3)^{2}},~~~~~k_2=\frac{-2n}{2n+3},~~~~~ 
\omega_0=\frac{3(8n^2+24n+17)}{(2n+3)^2}.\nonumber
\end{eqnarray} 
Substituting (17) and (19) into (15) and using (13) and (11), one gets
\begin{eqnarray}
\alpha (\phi,a) =-\frac{1}{3}{\beta_0} m \phi_0^{-1} a^n
(\phi/\phi_0)^{{l_1}-1}[(\phi/\phi_0)^{k_2}-\xi]^{l_2}[(\phi/\phi_0)^{k_2}+1]^{l_3}
\end{eqnarray}  
and 
\begin{eqnarray}
\beta (\phi,a) =\beta_0
a^n{(\phi/\phi_0)^{l_1}}[(\phi/\phi_0)^{k_2}-\xi]^{l_2}[(\phi/\phi_0)^{k_2}+1]^{l_3},
\end{eqnarray}
where $\beta_0$ is an integration constant, and we have used the definitions
\begin{eqnarray}
&{}&{l_1=\frac{m(\omega_0-k_1-1)}{2\omega_0-2k_1-m}} \nonumber\\
&{}&{l_2=\frac{2{k_1}^2k_2-k_2(m-2\omega_0)^2/2-2m(m-2)}{4{k_1}^2-(m-2\omega_0)^2}}\nonumber\\
&{}&{l_3=-\frac{2{k_1}^2k_2-k_2(m-2\omega_0)^2/2
-2m(m-2)^2(k_1+1) }{4{k_1}^2-(m-2\omega_0)^2}}\nonumber\\
&{}&{\xi=\frac{2k_1-(m-2\omega_0)}{2k_1+(m-2\omega_0)}}.\nonumber
\end{eqnarray}
Therefore, the vector field $X$ exists, and the existence of the Noether
symmetry leads to the determination of the class of potentials (16) and the dynamical
coupling function (19), for arbitrary values of $n$, except $n=-3/2$.
In principle, the existence of the Noether symmetry means that there must
exist a constant of motion. In this case one may compute it using the
Cartan one-form associated with the Lagrangian (3)
\begin{eqnarray}
\theta_{\cal{L}} \equiv \frac{ \partial {\cal{L}}}{\partial \dot{a}}da+
\frac{ \partial {\cal{L}}}{\partial \dot{\phi}}d\phi.\nonumber
\end{eqnarray}
Contracting $\theta_{\cal{L}}$ with $X$ gives the following required
constant of motion
\begin{eqnarray}
i_{X} \theta_{\cal{L}}=12a \phi\alpha(\phi,a) \{a\dot{\phi}+\phi\dot{a}\}
+4a^2 \beta(\phi,a) \{3\dot{a}\phi+2a \dot{\phi}\omega(\phi)\} \nonumber
\end{eqnarray}
where  $\omega(\phi)$, $\alpha(\phi,a)$ and $\beta(\phi,a)$ are given by
(19), (20) and (21), respectively.
\section {Dynamical Field Equations and Solutions}
Scalar-tensor theory reduces to Brans-Dicke theory
when the coupling function $\omega(\phi)$ is taken to be a constant.
For mathematical simplicity, we analyze the solutions of the field equations
(4) and (5) for $n=-3$ and in the case that $\omega(\phi)$ is a constant 
parameter such as $\omega$. In this case the Lagrangian (3) takes the form
\begin{eqnarray}
{\cal{L}}= 12\dot{a}a^2 \phi \dot{\phi}+ 6\dot{a}^2 a \phi^2+a^3
(4\omega\dot{\phi}^2-\lambda \phi^4).
\end{eqnarray}
The corresponding field equations are given by
\begin{eqnarray}
[\frac{\ddot{a}}{a}+\frac{1}{2}
(\frac{\dot{a}}{a})^2]\phi^2+[\ddot{\phi}+2(\frac{\dot{a}}{a})\dot{\phi}]\phi+{\dot{\phi}}^2
+\frac{1}{4}\lambda \phi^4=0
\end{eqnarray}
and
\begin{eqnarray}
\omega\ddot{\phi}+3\omega\dot{\phi}(\frac{\dot{a}}{a})+\frac{3}{2}
[\frac{\ddot{a}}{a}+(\frac{\dot{a}}{a})^2]\phi +\frac{1}{2}\lambda \phi^3=0.
\end{eqnarray}
One may select the initial conditions of the field equations (23) and (24)
such that the energy function associated with the Lagrangian (22) vanishes
\begin{eqnarray}
E_{\cal{L}}&=&{\frac{\partial {\cal{L}}}{\partial \dot{a}}\dot{a}+
\frac{\partial {\cal{L}}}{\partial\dot{\phi}}\dot{\phi}-{\cal{L}}}\nonumber\\ 
&=&{ 12\dot{a}a^2 \phi \dot{\phi}+ 6\dot{a}^2 a \phi^2+a^3 (4\omega
{ \dot{\phi}}^2+\lambda \phi^4)=0}.
\end{eqnarray}
In fact, by choosing $E_{\cal{L}}=0$, one imposes a constraint on the
integration constants. From (20) and (21) we have 
\begin{eqnarray}
\alpha=-\frac{4}{3} \frac{\beta_0}{a^2\phi^2}, 
~~~~~~\beta=\frac{\beta_0}{a^3 \phi}
\end{eqnarray}
for $n=-3$. Now, the existence of the vector field $X$ can be used to find
a non-degenerate point transformation
\begin{eqnarray}
 (a, \phi) \longrightarrow (p, q),\nonumber
\end{eqnarray}
such that the transformed Lagrangian is cyclic in one of the new
variables. A possible way is to compute the Cartan one-form associated
with Lagrangian $\cal{L}$, namely 
\begin{eqnarray}
\left\{ \matrix{ i_X(dp)=1 \nonumber \\i_X(dq)=0,}\right. \nonumber
\end{eqnarray}
or explicitly
\begin{eqnarray}
\left\{ \matrix{\alpha \frac{\partial p}{\partial a}+\beta
\frac{\partial p}{\partial \phi}=1 \nonumber \\ \alpha 
\frac{\partial q}{\partial a}+\beta \frac{\partial q}{\partial \phi}=0.}\right.
\end{eqnarray}  
Substituting (26) into (27), yields the solutions
\begin{eqnarray}
\left\{\matrix {p=-\frac{1}{2\beta_0}a^3\phi^2 \nonumber \\
q=\gamma_0 a^{3/\gamma} \phi^{4/\gamma},}\right.
\end{eqnarray}
where $\gamma_0$ and $\gamma$ are constant.
Under these transformations, the Lagrangian (22) takes the form
\begin{eqnarray}
{\cal{L}}=2\beta_0 \left \{2\gamma(\omega-5/3)
\frac{\dot{p}\dot{q}}{q}+(\omega-4/3)
(\frac{{\dot{p}}^2}{p}+\gamma^2 \frac{p{\dot{q}}^2}{q^2})+
\frac{\lambda}{2\beta_0 \gamma_0}q^{\gamma}\right \}.
\end{eqnarray}
One can easily check that Eq. (18) has two solutions: $\omega=3/2$ and
$\omega=4/3$ for $n=-3$. Now, if one computes the
Hessian determinant ${\cal{H}} 
=|{\partial^2 {\cal{L}}}/{\partial \dot{\phi}}{\partial \dot{a}}|$, 
one finds that  $\omega=3/2$ leads to the degeneration of $\cal{L}$. 
In other words, this value of $\omega$ does not allow a Hamiltonian
formulation for the theory and leads to pathological dynamics. Thus,
one can disregard $\omega =3/2$. For $\omega=4/3$ the second term of
(29) disappears and $\cal{L}$ takes the form
\begin{eqnarray}
{\cal{L}}=-\frac{4}{3} \beta_0 \gamma\frac{\dot{p}\dot{q}}{q}+
\frac{\lambda}{\gamma_0}q^{\gamma}.
\end{eqnarray}
This Lagrangian clearly does not depend on $p$. Therefore, in new
configuration space $(p,q)$, the variable $p$ is cyclic.
This implies the existence of the Noether symmetry. \\
The corresponding field equations for the last Lagrangian are given by 
\begin{eqnarray}
\ddot{p}+\frac{3\lambda}{4\beta_0 \gamma_0}q^{\gamma}=0\nonumber
\end{eqnarray}
and
\begin{eqnarray}
\dot{q}-r_0 q=0\nonumber
\end{eqnarray}
where $r_0$ is constant of motion. The solutions of these equations are  
\begin{eqnarray}
p=s_0(e^{r_0\gamma t}+p_0)+\dot{{p_0}}t
\end{eqnarray}
and
\begin{eqnarray}
q=q_0 e^{r_0 t}, 
\end{eqnarray}
where $p_0, \dot{{p_0}}$ and $q_0$ are integration constants and $s_0$ 
is given by
\begin{eqnarray}
s_0=-\frac{3\lambda {q_0}^{\gamma}}{4\beta_0 \gamma_0 {\gamma}^2
{r_0}^2}.\nonumber
\end{eqnarray}
In new configuration space $(p, q)$, the condition (25) can be rewritten as 
\begin{eqnarray}
E_{\cal{L}}=-\frac{4}{3} \beta_0 \gamma\frac{\dot{p}\dot{q}}{q}-
\frac{\lambda}{\gamma_0}q^{\gamma}=0.
\end{eqnarray}
Substituting (31) and (32) into (33), leads to $\dot{{p_0}}=0$, and so the
solution (31) reduces to
\begin{eqnarray}
p=s_0(e^{r_0\gamma t}+p_0).
\end{eqnarray}
Putting (32) and (34) into (28), one gets the following solutions 
\begin{eqnarray}
\phi=\psi_0e^{r_0\gamma t/2}(e^{r_0\gamma t}+p_0)^{-1/2},~~~~\psi_0 
=(-\frac{{q_0}^\gamma}{2s_0\beta_0 \gamma_0^\gamma})^{1/2}\nonumber
\end{eqnarray}
and 
\begin{eqnarray}
a=a_0e^{-r_0\gamma t}(e^{r_0\gamma t}+p_0)^{2/3},~~~~a_0
=(\frac{4s_0^2 \beta_0^2 \gamma_0^\gamma}{q_0^\gamma})^{1/3}. \nonumber
\end{eqnarray}
In the limit $t \rightarrow \infty$ the scalar field $\phi$
approaches a constant value $\psi_0$, and then the Einstein 
gravity is recovered and the Newtonian gravitational constant is 
identified as $G_N=\frac{1}{\psi_0}^2$. 
\section{Concluding Remarks}
We have studied scalar-tensor theories of gravity in which the coupling 
function $\omega(\phi)$ and the generic potential $V(\phi)$ are unknown. 
The form of the coupling function and the potential are  determined
using Noether symmetry in a flat FRW background. 
In the special case $n=-3$, we showed that Noether symmetry exists for
$\omega=3/4$, and we derived the corresponding constant of motion, $r_0$.
Furthermore, the exact solutions of the field equations were derived from a
point-like Lagrangian for a quartic potential.
It is also shown that in the case of $\omega=3/2$, there is no Noether
symmetry for the Lagrangian. However, it is interesting to note that the
theory is exactly conformal invariant.${}^{9}$ Further attempts seems to be
necessary to understand this relationship more deeply.\\


\begin{thebibliography}{99}
\bibitem{1} A. Zee, Phys. Rev. Lett. {\bf 42}, 417 (1979);  \\
            L. Smolin, Nucl. Phys. {\bf B 160}, 253 (1979);       \\
            S. Adler, Phys. Rev. Lett. {\bf 44}, 1567 (1980).
\bibitem{2} S. Capozziello and R. De Ritis Phys. Lett. {\bf A177}, 1
            (1993).
\bibitem{3} P. Jordan, Z. Phys. {\bf 157,} 112 (1959).
\bibitem{4} C. Brans and R. H. Dicke, Phys. Rev. {\bf 124,} 925 (1961).
\bibitem{5} P. Teyssandier and P. Tourrence, J. Math. Phys. {\bf 24,} 2793 
            (1983);\\ D. Wands, Class. Quantum Grav. {\bf 7,} 269 (1994).
   
\bibitem{6} R. De Ritis, G. Platania, P. Scudellaro, and C. Stornaiolo,
            Gen. Relativ. Gravit. {\bf 22}, 97 (1990).
\bibitem{7} S. Capozziello, R. De Ritis, C. Rubano and P. Scudellaro
            Riv, Nuovo Cimento {\bf 4} (1996).
\bibitem{8} R. Abraham and J. Marsden 1978 Foundation of Mechanics (New
            York; Benjamin) \\ 
            G. Marmo, E. J. Saletan, A. Simoni and B. Vitale 1985
            Dynamical Systems.\\ A Differential Geometric Approach to
            Symmetry and Reduction (New York: Wiley). 
\bibitem{9} H. Motavali, H. Salehi and M. Golshani,  
            Int. J. Mod. Phys. {\bf A15}, 983 (2000).\\
            H. Motavali, H. Salehi and M. Golshani,             
            Int. J. Mod. Phys. Lett. {\bf A14}, 2481 (1999).
\end{thebibliography}
\end{document}